\documentstyle[aps,pra,epsf]{revtex}

\begin{document}
\title{Critical temperature and Ginzburg-Landau equation for a trapped Fermi gas.}
\author{M.A. Baranov and D.S. Petrov}
\address{Russian Research Center ''Kurchatov Institute'',\\
Kurchatov Square, 123182 Moscow, Russia}
\date{December 19, 1997}
\maketitle

\begin{abstract}
\tightenlines
We discuss a superfluid phase transition in a trapped neutral-atom Fermi
gas. We consider the case where the critical temperature greatly exceeds the
spacing between the trap levels and derive the corresponding Ginzburg-Landau
equation. The latter turns out to be analogous to the equation for the
condensate wave function in a trapped Bose gas. The analysis of its solution
provides us with the value of the critical temperature $T_{c}$ and with the
spatial and temperature dependence of the order parameter in the vicinity of
the phase transition point.
\end{abstract}

\vspace{4mm}
The recent progress in the studies of ultra-cold trapped Bose gases and the
discovery of Bose-Einstein condensation \cite{Cor95,Hul95,Ket95} stimulate
an interest to macroscopic quantum phenomena in trapped Fermi gases. The
most prominent phenomena should be connected with a superfluid phase
transition associated with the appearance of the order parameter --
macroscopic wave function of strongly correlated two-particle states on the
Fermi surface (Cooper pairs). The possibilities of finding this phase
transition in trapped Fermi gases have been discussed in Refs.
\cite{HS,BKK,HS1}.

A remarkable feature of neutral-atom Fermi gases is that the Cooper pairing
and, hence, the superfluid phase transition can occur for both attractive
and repulsive interparticle interaction. For an attractive interaction
(negative scattering length $a$) the pairing occurs in the $s$-wave channel,
as described by the standard BCS approach, and one has ''singlet'' Cooper
pairs. In a neutral-atom Fermi gas, according to the Pauli principle, such a
pair can only be formed by two atoms which are in different hyperfine
states. Therefore, the critical temperature $T_{c}$ of the transition is
very sensitive to the difference in concentrations of the two hyperfine
components, and under the condition $\Delta n/n\equiv
(n_{1}-n_{2})/(n_{1}+n_{2})\gtrsim T_{c}/\varepsilon _{F}\ll 1$, where $%
\varepsilon _{F}\sim \hbar ^{2}n^{2/3}/m$ is the Fermi energy, there will be
a complete suppression of the spin-singlet pairing. For $^{6}$Li with $%
a\approx -1140$\AA \cite{H}, one has $T_{c}\approx 30{\rm nK}$ for the atom
density $n=4\cdot 10^{12}{\rm cm}^{-3}$ \cite{HS}, and the existence of the $%
s$-wave pairing requires $\Delta n/n<10^{-2}$.

For positive scattering length (repulsive interaction) the $s$-wave pairing
is impossible, and one has to consider the mechanism of $p$-wave ''triplet''
pairing which originates from the effective interaction caused by
polarization effects \cite{BCK}. Actually, this pairing mechanism is
insensitive to the sign of the scattering length. It works equally well for $%
a<0$ in the situation where the direct $s$-wave pairing is suppressed. In
these cases the Pauli principle allows to have Cooper pairs formed by two
particles which are in one and the same hyperfine state, whereas the
particles in other hyperfine states participate only in the formation of the
effective pairing interaction. Therefore, the $p$-wave "triplet" pairing
does not require any severe restriction on $\Delta n$. As found \cite{KC},
the corresponding critical temperature $T_{c}$ depends non-monotonically on $%
n_1$, $n_2$ and becomes zero only in the case where all particles are in the
same hyperfine state. Since the effective interaction based on polarization
effects is weaker than the direct interparticle interaction, the $p$-wave
"triplet" pairing results in a lower value of $T_{c}$ as compared to the
critical temperature for the $s$-wave pairing. For example, for the $p$-wave
pairing in a gas of $^6Li$ the critical temperature $T_{c}\approx 30{\rm nK}$
corresponds to densities $n\approx 10^{13}cm^{-3}$.

In this paper we study the influence of a (harmonic) trapping potential on
superfluid pairing. We derive the corresponding Ginzburg-Landau (GL)
equation for the order parameter, assuming the critical temperature $T_{c}$
much higher than the level spacing $\Omega$ in the trap. The analysis of
this equation provides us with the value of $T_{c}$ for the trapped gas and
gives the coordinate and temperature dependence of the order parameter in
the vicinity of the phase transition. As found, the critical temperature is
slightly lower than that for a spatially homogeneous Fermi gas with density $%
n_0$ (maximum density of the trapped gas), and the behavior of the order
parameter resembles the behavior of a trapped Bose condensate.

We consider a two-component neutral gas of fermionic atoms, with a
short-range interatomic interaction, trapped in a spherically symmetric
harmonic potential. The two (hyperfine) components are labeled by indices $%
\alpha =\pm $ and are assumed to have equal concentrations. The Hamiltonian
of the system has the form ($\hbar =1$)
\begin{equation}
H=\int_{{\bf r}}\sum_{\alpha }\psi _{\alpha }^{+}({\bf r)}\left( -\frac{1}{2m%
}\partial ^{2}-\mu +\frac{m\Omega ^{2}{\bf r}^{2}}{2}\right) \psi _{\alpha }(%
{\bf r)+}\frac{g}{2}\sum_{\alpha ,\beta }\int_{{\bf r}}\psi _{\alpha }^{+}(%
{\bf r)}\psi _{\alpha }({\bf r)}\psi _{\beta }^{+}({\bf r)}\psi _{\beta }(%
{\bf r}),  \label{1}
\end{equation}
where $\mu $ is the chemical potential (Fermi energy), $\Omega $ the trap
frequency, $g=4\pi a/m$ the interaction strength, and $m$ the atom mass.

The interaction effects can be expressed in terms of a small gaseous
parameter $\lambda =2|a|p_{F}/\pi <1$, where $p_{F}=mv_F=(3%
\pi^{2}n_{0})^{1/3}$ is the Fermi momentum. In the spatially homogeneous
case the system of fermions described by the Hamiltonian (\ref{1}) with $%
\Omega =0$, undergoes a superfluid phase transition. The transition
temperature $T_c^{(0)}$ and the type of pairing depend on the sign of $a$.
For negative $a$ (attractive interaction) there will be the $s$-wave
"singlet" pairing, whereas for positive $a$ the $p$-wave ''triplet'' pairing
should take place (see Ref. \cite{BCK} for details).
In both cases the critical
temperature $T_{c}^{(0)}=C\varepsilon _{F}\exp \{-1/\Gamma \}$, where $%
\varepsilon _{F}=p_{F}^{2}/2m$ is the Fermi energy, $\Gamma$ the pairing
interaction, and $C$ a numerical coefficient of order unity. The value of $C$
and the expression for $\Gamma $ depend on the type of pairing. For the
''singlet'' pairing one has $\Gamma =\lambda$, and for the "triplet" pairing
$\Gamma\approx\lambda^{2}/13$.

In the Thomas-Fermi approach ($\varepsilon _{F}\gg \Omega $) the density
profile of the trapped Fermi gas is $n(r)=n_{0}(1-(r/R_{TF})^{2})^{3/2}$,
and the Thomas-Fermi radius $R_{TF}=v_{F}/\Omega $ turns out to be the
natural length scale in the system. One can also introduce the local Fermi
momentum $p_{F}(r)=p_{F}(1-(r/R_{TF})^{2})^{1/2}$ and the density of states
on the local Fermi surface, $\nu _{0}(r)=mp_{F}(r)/(2\pi ^{2})$. It should
be noted that all these quantities are only slightly influenced by the
superfluid pairing, because the latter involves only a small fraction of
particles ($\sim T_{c}/\varepsilon _{F}\ll 1$), with energies close to the
Fermi energy.

For describing the phase transition one has to introduce the order parameter
which in the case of ''singlet'' pairing is a complex function $\Delta ({\bf %
r})\sim \left\langle \psi _{\sigma }({\bf r})\psi _{\sigma ^{\prime }}({\bf r%
})\right\rangle \varepsilon _{\sigma \sigma ^{\prime }}$, with $\varepsilon
_{\sigma \sigma ^{\prime }}$ being the antisymmetric tensor. For the
''triplet'' pairing the order parameter is a $3\times 3$ complex matrix $%
\Delta _{ij}({\bf r})\sim \left( \sigma _{2}\sigma _{i}\right) _{\alpha
\beta }\left\langle \psi _{\alpha }({\bf r})\partial _{j}\psi _{\beta }({\bf %
r})\right\rangle $, where $\sigma _{i}$ are the Pauli matrices. The
time-independent Ginzburg-Landau (GL) equation describes the equilibrium
behavior of the order parameter $\Delta ({\bf r)}$ below $T_{c}$, assuming $%
T_{c}-T\ll T_{c}$. The critical temperature $T_{c}$ can be found as
temperature below which this equation has a nontrivial solution. We present
the derivation of the GL equation for the trapped Fermi gas in the case of
''singlet'' pairing ($g<0$), relying on the assumption that
\[
T_{c},T_{c}^{(0)}\gg \Omega .
\]
The derivation of the GL equation for the ''triplet'' pairing can be
performed along the same lines, and should be based on the results of Ref.
\cite{BK}.

For the $s$-wave ''singlet'' pairing the equilibrium GL free energy ($\Delta
$-dependent part of the free energy) can be written in the form
\begin{equation}
F_{GL}[\Delta ]=\int_{{\bf r}}\frac{\left| \Delta ({\bf r})\right| ^{2}}{%
\left| g\right| }-T\ln \left\langle T_{\tau }exp\left\{
-\int\limits_{0}^{1/T}d\tau \int_{{\bf r}}\left( \psi _{+}({\bf r},\tau
)\psi _{-}({\bf r},\tau )\Delta ^{*}({\bf r})+h.c.\right) \right\}
\right\rangle _{0},  \label{2}
\end{equation}
where we use the Matsubara representation, and the symbol $<\ldots >_{0}$
stands for the average over the states of the free-particle Hamiltonian
(first term in Eq. (\ref{1})). In the vicinity of the phase transition the
quantity $\Delta $ is small, and the second term in Eq.(\ref{2}) can be
expanded in powers of $\Delta $. As usual, we perform this expansion up to
the fourth power. The coefficients of the expansions (kernels) can be
expressed in terms of Green function of the normal state (without pairing), $%
G_{\omega }^{(0)}({\bf r}_{1},{\bf r}_{2})$. As will be justified below, the
order parameter varies on a distance scale $l_{\Delta }$ which is much
larger than the characteristic distance scale $\xi _{K}\sim v_{F}/T_{c}$ of
these kernels. Accordingly, $F_{GL}[\Delta ]$ can be represented as
(hereinafter we use $R_{TF}$ as a unit of length)
\begin{equation}
F_{GL}\left[ \Delta \right] =R_{TF}^{3}\int_{{\bf R}}\left\{ \frac{\left|
\Delta \right| ^{2}}{\left| g\right| }-K_{0}^{(2)}({\bf R)|}\Delta {\bf |}%
^{2}-K_{1}^{(2)}({\bf R)}\left( \Delta \partial ^{2}\Delta ^{*}+\Delta
^{*}\partial ^{2}\Delta -2\partial _{i}\Delta ^{*}\,\partial _{i}\Delta
\right) +K^{(4)}{\bf |}\Delta {\bf |}^{4}\right\} {\bf ,}  \label{4}
\end{equation}
where ${\bf R}=({\bf r}_{1}+{\bf r}_{2})/2$, ${\bf r}={\bf r}_{1}-{\bf r}%
_{2} $, $\Delta \equiv \Delta ({\bf R})$, and

\begin{equation}
K^{(4)}({\bf R)=}\nu _{0}({\bf R})\frac{7\zeta (3)}{16\pi ^{2}T^{2}},
\label{6}
\end{equation}
\begin{eqnarray}
K_{0}^{(2)}({\bf R)} &=&R_{TF}^{3}T\sum\limits_{\omega }\int_{{\bf r}%
}G_{-\omega }^{(0)}({\bf R},{\bf r})G_{\omega }^{(0)}({\bf R},{\bf r}),\quad
\label{5} \\
K_{1}^{(2)}({\bf R})\delta _{ij} &=&R_{TF}^{5}\frac{T}{8}\sum\limits_{\omega
}\int_{{\bf r}}r_{i}r_{j}G_{-\omega }^{(0)}({\bf R},{\bf r})G_{\omega
}^{(0)}({\bf R},{\bf r}).
\end{eqnarray}
Here $\zeta (x)$ is the Riemann zeta-function, and the summation is
performed over the Matsubara frequencies $\omega =\pi T(2n+1)$, $n=0,\pm
1,...$. The condition $T_{c}^{(0)}\gg \Omega $ allows us to use the
quasiclassical expression for the product of two Green functions:
\begin{equation}
G_{-\omega }^{(0)}({\bf R},{\bf r})G_{\omega }^{(0)}({\bf R},{\bf r})=\left(
\frac{m}{2\pi rR_{TF}}\right) ^{2}\exp \left\{ -r\frac{\left| \omega \right|
}{\Omega }\frac{2\sqrt{2}}{\left[ \sqrt{\left( 1-R^{2}\right) ^{2}+\left(
\omega /\varepsilon _{F}\right) ^{2}}+1-R^{2}\right] ^{1/2}}\right\} ,\quad
\label{7}
\end{equation}
which can be obtained from the corresponding expression for the spatially
homogeneous case, with the replacement $p_{F}\rightarrow p_{F}(R)$. The
validity of Eq. (\ref{7}) requires the condition $(1-R^{2})\gg (\Omega
/T)^{2}$. The use of Eq. (\ref{7}) for calculating the kernels $K_{0}^{(2)}$
and $K_{1}^{2}$ is justified by the fact that the pairing takes place only
in the central region of the gas cloud, and the characteristic size of this
region $l_{\Delta }\ll 1$. The main contribution to $K_{1}^{(2)}$ comes from
frequencies $\left| \omega \right| \ll \varepsilon _{F}$, and a
straightforward calculation yields
\begin{equation}
K_{1}^{(2)}({\bf R)=}\frac{1}{4}\nu _{0}(R)\kappa ^{2},  \label{8}
\end{equation}
where $\kappa =\sqrt{7\varsigma (3)/48\pi ^{2}}\left( \Omega /T\right)
=0.13\cdot \left( \Omega /T\right) \ll 1$.

The calculation of $K_{0}^{(2)}$ is more subtle, because the frequency sum
in Eq. (\ref{5}) diverges. The divergency can be eliminated in a standard
way by renormalization of the bare interaction $g$, and finally one has
\[
\frac{1}{\left| g\right| }-K_{0}^{(2)}({\bf R)=}\frac{1}{\left| a\right| }%
-\nu _{0}(R)\ln \frac{C\varepsilon _{F}(R)}{T}.
\]

Then, the final expression for the GL free energy can be written in the form
\begin{equation}
F_{GL}\left[ \Delta \right] =R_{TF}^{3}\int_{{\bf R}}\nu _{0}(R)\left\{ -%
\frac{\kappa ^{2}}{4}\left( \ldots \right) +\left[ \frac{1}{\lambda }\left(
\frac{\nu _{0}}{\nu _{0}(R)}-1\right) -\ln \left( \frac{T_{c}^{(0)}}{T}\frac{%
\varepsilon _{F}(R)}{\varepsilon _{F}}\right) \right] \left| \Delta \right|
^{2}+\frac{7\zeta (3)}{16\pi ^{2}}\frac{1}{T^{2}}\left| \Delta \right|
^{4}\right\} ,  \label{9}
\end{equation}
where $\nu _{0}(R)=\nu _{0}\sqrt{1-R^{2}}$ , $\varepsilon
_{F}(R)=\varepsilon _{F}\cdot (1-R^{2})$, and the symbol $\left( \ldots
\right) $ stands for the same combination of derivatives of $\Delta $ as in
Eq. (\ref{4}). The analogous expression for the ''triplet'' pairing can be
obtained from Eq. (\ref{9}), with the replacements $\lambda \rightarrow
\Gamma \approx \lambda ^{2}/13$, $\left| \Delta \right| ^{2}\rightarrow {\rm %
tr}\left( \Delta ^{+}\Delta \right) $, and $\left| \Delta \right|
^{4}\rightarrow {\rm tr}\left( \Delta ^{+}\Delta \right) ^{2}$.

Since only small distances $R\sim \sqrt{\kappa }\ll 1$ in Eq. (\ref{9}) are
important for pairing, we will make an expansion in powers of $R$ and retain
only the largest (quadratic) terms. Then the minimization of Eq. (\ref{9})
with respect to $\Delta ^{*}$ gives the GL equation:
\begin{equation}
\left[ -\kappa ^{2}\partial ^{2}+\left( \frac{1+2\lambda }{2\lambda }\right)
R^{2}-\ln \left( \frac{T_{c}^{(0)}}{T}\right) \right] \Delta +\frac{7\zeta
(3)}{8\pi ^{2}}\frac{\left| \Delta \right| ^{2}}{T^{2}}\Delta =0.  \label{10}
\end{equation}
We stress once more that Eq.(\ref{10}) is valid under the condition $\Delta
/T\ll 1$, which, in turn, implies that $T_{c}-T\ll T_{c}$.

It is interesting to emphasize that Eq. (\ref{10}) for $\Delta $ is formally
equivalent to the non-linear Schr\"{o}dinger equation for the condensate
wave function $\Psi _{0}$ in a Bose gas of neutral particles of "mass" $%
1/2\kappa^2$ in a harmonic confining potential with "frequency" ("level
spacing") $2\widetilde{\kappa}=2\kappa(1+1/2\lambda)^{1/2}$. The last
(non-linear) term in the l.h.s. plays a role of repulsive interparticle
interaction, and the third term $\ln(T_c^{(0)}/T)$ the role of the chemical
potential. Accordingly, the calculation of the shape of $\Delta$ is similar
to the calculation of the shape of $\Psi_0$ in a trapped Bose gas, performed
in e.g. \cite{EB,KSW}. But there is an important difference. In the Bose gas
the amplitude of $\Psi _{0}$ is determined by the normalization condition.
Together with the Schr\"{o}dinger equation this condition gives the chemical
potential as a function of the particle number. Hence, for small
interparticle interaction the non-linear term is not important at all. In
the Fermi gas the amplitude of $\Delta $ is always determined by the
non-linear term.

The critical temperature $T_{c}$ for the trapped Fermi gas is the maximum
temperature $T$ at which Eq.(\ref{10}) has a nontrivial solution. As usual, $%
\Delta \rightarrow 0$ for $T\rightarrow T_{c}$, and for finding $T_{c}$ the
non-linear term in Eq. (\ref{10}) can be omitted. Then the GL equation
becomes similar to the Schr\"{o}dinger equation for spherically symmetrical
oscillator, and we obtain
\begin{equation}
\frac{T_{c}^{(0)}-T_{c}}{T_{c}^{(0)}}\approx \ln \frac{T_{c}^{(0)}}{T_{c}}=3%
\widetilde{\kappa }\ll 1.  \label{11}
\end{equation}
One can see from Eq. (\ref{11}) that the critical temperature $T_{c}$ for
the trapped gas is only slightly lower than $T_{c}^{(0)}$ for the
homogeneous gas with density $n_{0} $.

As well as in the case of a trapped Bose condensate, the shape of the order
parameter in Eq.(\ref{10}) is predetermined by the ratio $z$ of the
non-linear term $|\Delta|^2/T^2$ to the "level spacing" $2\widetilde{\kappa}$%
. Since the non-linear term is of order the difference between the "chemical
potential" $\ln(T_c^{(0)}/T)$ and its minimum value (\ref{11}), we have
\begin{equation}
z=\frac{1}{2\widetilde{\kappa}}\ln\left( \frac{T_c}{T}\right) \approx\frac{%
\delta T}{\Omega}(1+1/2\lambda)^{-1/2},  \label{z}
\end{equation}
where $\delta T=T_c-T$.

For $z\ll 1$ and, hence, $T$ very close to $T_{c}$, the non-linear term in
Eq.(\ref{10}) does not influence the shape of the order parameter, and the
latter takes the form of a Gaussian:
\begin{equation}
\Delta _{c}({\bf R})\sim \varphi _{0}\left( R\right) \equiv (\pi l_{\Delta
}^{2})^{-3/4}exp\left( -R^{2}/2l_{\Delta }^{2}\right) .  \label{12}
\end{equation}
The linear size of the spatial region where the pairing takes place, $%
l_{\Delta }=\kappa /\sqrt{\widetilde{\kappa }}\ll 1$, is finite for $T\to
T_{c}$. Moreover, $l_{\Delta }\gg \xi _{K}\sim \Omega /T_{c}$, which
justifies the gradient expansion in Eq. (\ref{4}). For finding the amplitude
of the order parameter, $\Delta (R=0)$, in the limiting
case $z\ll 1$ we write
$\Delta $ in the form $\Delta \left( {\bf R,}T\right) =\alpha \left(
T\right) \,(\varphi _{0}\left( R\right) +\delta \varphi ({\bf R},T))$, where
$\,\varphi =\varphi _{0}+\delta \varphi $ obeys the normalization condition $%
\int_{{\bf R}}\left| \varphi \left( {\bf R},T\right) \right| ^{2}{\bf =}1$,
and $\delta \varphi \rightarrow 0$ for $T\rightarrow T_{c}$. Then Eq. (\ref
{10}) is transformed to
\begin{equation}
\left[ \kappa ^{2}\partial ^{2}-\left( \frac{1+2\lambda }{2\lambda }\right)
R^{2}+3\widetilde{\kappa }\right] \delta \varphi +\ln \left( T_{c}/T\right)
(\varphi _{0}+\delta \varphi )=\alpha ^{2}\frac{7\zeta (3)}{8\pi ^{2}}\frac{%
(\varphi _{0}+\delta \varphi )^{3}}{T_{c}^{2}},  \label{14}
\end{equation}
Eq. (\ref{14}) gives $\alpha $ and $\delta \varphi $ as series of rational
powers of $\ln (T_{c}/T)\approx \delta T/T_{c}$, the small parameter of
expansion being $z$. Multiplying both sides of this equation by $\varphi
_{0}\left( {\bf R}\right) $, integrating over ${\bf R}$, and omitting the
terms containing $\delta \varphi $, to the leading order we obtain
\[
\alpha =T_{c}l_{\Delta }^{3/2}\left( \frac{16\pi ^{3}\sqrt{2\pi }}{7\zeta (3)%
}\ln {\frac{T_{c}}{T}}\right) ^{1/2},
\]
and, hence,
\begin{equation}
\Delta \left( {\bf R},T\right) \approx \Delta _{0}\left( {\bf R},T\right)
=T_{c}\sqrt{\frac{16\pi ^{2}\sqrt{2}}{7\zeta (3)}\ln \frac{T_{c}}{T}}\cdot
exp\left( -\frac{R^{2}}{2l_{\Delta }^{2}}\right) \approx 5.15T_{c}\sqrt{%
\frac{T_{c}-T}{T_{c}}}exp\left( -\frac{R^{2}}{2l_{\Delta }^{2}}\right) .
\label{16}
\end{equation}
As well as in the spatially homogeneous case, we have $\Delta \sim \sqrt{%
T_{c}-T}$ for $T\to T_{c}$. For the ''triplet'' pairing one will have $%
\Delta _{ij}({\bf R},T)=\delta _{ij}\Delta _{0}\left( {\bf R,}T\right) $.
It is important to mention, however, that the corrections to Eq.(\ref{16}),
which can be obtained from Eq. (\ref{14}), are physically meaningless. They
have the same order of magnitude ($\propto (\delta T/T_{c})^{3/2}$) as the
corrections originating from terms with higher powers of $\Delta $ (for
example, $\Delta ^{5}/T_{c}^{4}$) or higher derivatives, neglected in
deriving Eq. (\ref{10}). For the same reason one should not go beyond the
first term in expanding $\ln (T_{c}/T)$ in powers of $\delta T/T_{c}$. In
Fig.1 we present the approximate solution $\Delta _{0}$ (solid lines) and
the corresponding numerical solutions of Eq. (\ref{10}) (dashed lines) for $%
T_{c}^{(0)}/\Omega =5$, $\lambda =0.3$, and $\delta T/T_{c}=0.001$, $0.01$, $%
0.03$. For these values of $T_{c}^{(0)}/\Omega$ and $\lambda$ we have $%
\widetilde{\kappa }=4.4\cdot 10^{-2}$, and Eq. (\ref{11}) gives the critical
temperature $T_{c}=0.87T_{c}^{(0)}$ which is only $1\%$ higher than $T_c$
following from the exact numerical solution of Eq.(\ref{10}).

\begin{figure}
\epsfysize=10cm
\centerline{\epsfbox{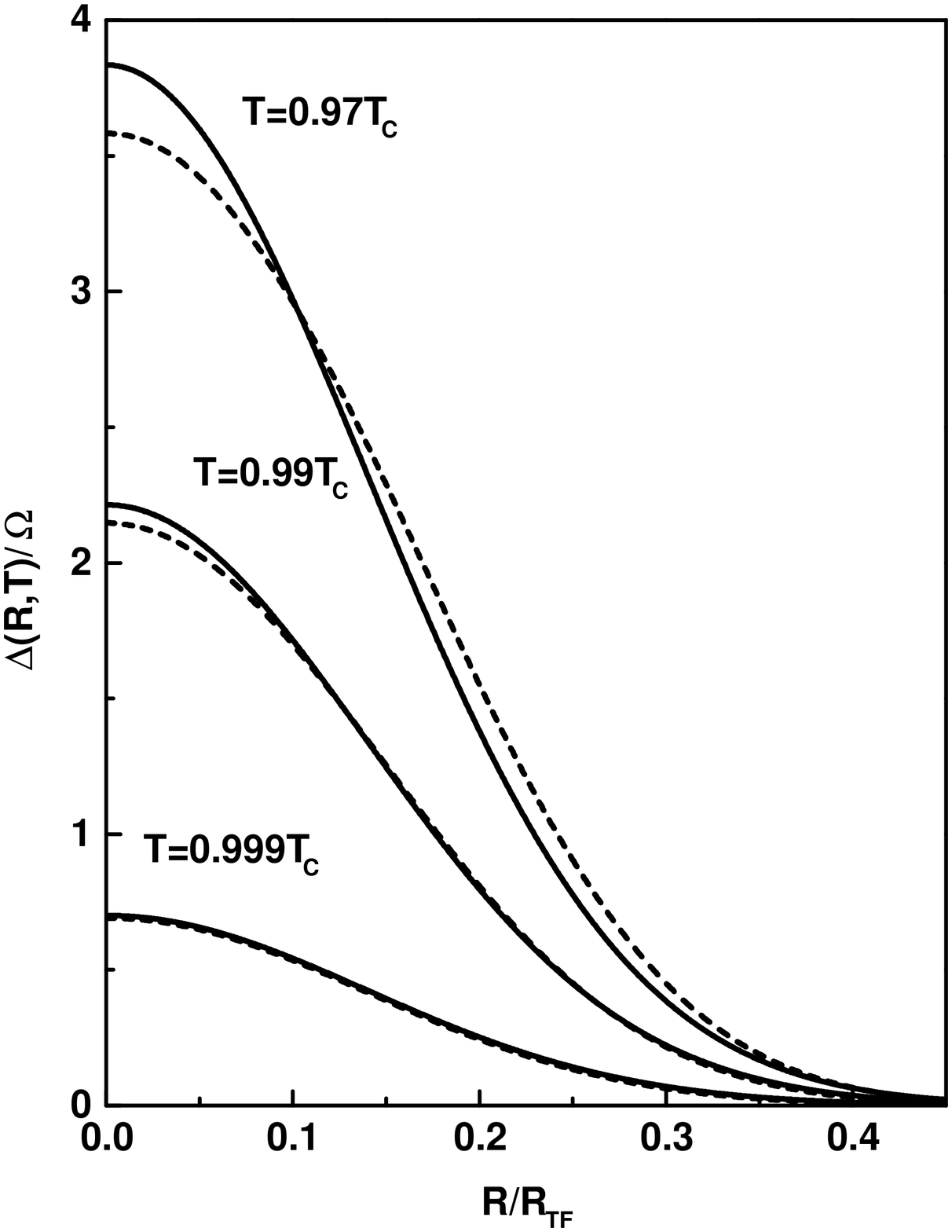}}
\caption{The order parameter versus $R$ for various temperatures. The solid
lines correspond to $\Delta _{0}(R,T)$ (15), and the dashed lines to
numerical solutions of Eq. (10). }
\label{f1}
\end{figure}
For lower temperature, where $z\gg 1$ (but still much smaller than $%
\widetilde{\kappa }^{-1}$, as required by the condition $\delta T\ll T_{c}$%
), one can neglect the Laplacian term in Eq. (\ref{10}), and write the
approximate solution for the order parameter in the form:
\begin{equation}
\Delta \left( {\bf R,}T\right) =T_{c}\sqrt{\frac{8\pi ^{2}}{7\zeta (3)}\ln {%
\frac{T_{c}}{T}}}\left( 1-R^{2}/R_{c}^{2}\right) ^{1/2}\approx 3.06T_{c}%
\sqrt{\frac{T_{c}-T}{T_{c}}}\left( 1-R^{2}/R_{c}^{2}\right) ^{1/2}
\label{18}
\end{equation}
for $R\leq R_{c}=\sqrt{(\delta T/T_{c}){(1+1/2\lambda )}^{-1}}=l_{\Delta}
\sqrt{2z}\ll 1$, and zero otherwise. The solution (\ref{18}) is completely
analogous to that for the Bose condensate wave function in the
quasiclassical (Thomas-Fermi) regime \cite{GSL,HuS}.

Eqs. (\ref{16}) and (\ref{18}) show that in the vicinity of the phase
transition the superfluid pairing takes place only in a small central region
of the gas sample. This, together with the fact that the superfluid pairing
involves only a small fraction ($\sim T_{c}/\varepsilon _{F}\ll 1$) of
atoms, makes it very difficult to detect the presence of pairing through the
measurement of the gas density profile. On the other hand, as well as in the
spatially homogeneous case, the pairing should influence the spectrum of
elementary excitations. In this respect we believe that the measurement of
eigenfrequencies of oscillations of the gas cloud can be one of the most
promising ways of identifying the phase transition in trapped Fermi gases.

We acknowledge very fruitful discussion with G.V. Shlyapnikov. This work was
supported by the Russian Foundation for Basic Studies (grant 97-02-16532).

\end{document}